# DEVELOPMENT OF SUPERCONDUCTING LINAC FOR THE KEK/JAERI JOINT PROJECT

M.Mizumoto, N.Ouchi, J.Kusano, E.Chishiro, K.Hasegawa, N.Akaoka, JAERI, Tokai, Japan
K.Saito, S.Noguchi, E.Kako, H.Inoue, T.Shishido, M.Ono, KEK, Tsukuba, Japan
K.Mukugi, C.Tsukishima, MELCO, Kobe, Japan
O.Takeda, Toshiba Corporation, Kawasaki, Japan
M.Matsuoka, MHI, Kobe, Japan

*Abstract*

The JAERI/KEK Joint Project for the high-intensity proton accelerator facility has been proposed with a superconducting (SC) linac option from 400 MeV to 600MeV. System design of the SC linac has been carried out based on the equipartitioning concept. The SC linac is planned to use as an injector to a 3GeV rapid cycling synchrotron (RCS) for spallation neutron source after it meets requirement to momentum spread less than ±0.1%. In the R&D work for SC cavities, vertical tests of single-cell and 5 cell cavities were performed. Experiments on multi-cell (5 cell) cavities of $\beta=0.50$ and $\beta=0.89$ at 2K were carried out with values of maximum electric surface peak fields of 23MV/m and 31MV/m, respectively. A model describing dynamic Lorentz detuning for SC cavities has been developed for pulse mode operation. Validity of the model was confirmed experimentally to simulate the performance.

## 1 INTRODUCTION

The Japan Atomic Energy Research Institute (JAERI) and the High Energy Accelerator Research Organization (KEK) are proposing the Joint Project for High Intensity Proton Accelerator[1] by merging their original Neutron Science Project (NSP)[2] and Japan Hadron Facility (JHF)[3]. The accelerator complex for the Joint Project consists of a 600-MeV linac, a 3-GeV RCS and a 50-GeV synchrotron. The linac comprises a negative ion source, a 3-MeV RFQ, a 50-MeV DTL, a 200-MeV SDTL (Separated type DTL), a 400-MeV CCL and a 600-MeV SC linac. Frequency of RFQ, DTL and SDTL is 324 MHz. Frequency of CCL and SC linac is 972 MHz. The 400MeV beams are injected to the RCS in the first step of the Project. Small momentum spread, $\Delta p/p$ less than ±0.1%, is required to inject to the RCS. The 600MeV SC linac will be used to improve the beam intensity after acceptable beams to the RCS be achieved in pulsed operation.

The R&D studies for the SC linac have been carried out at JAERI in collaboration with KEK. Dynamic behavior of the Lorentz detuning is important for stable RF control. Lorentz vibration model was established to describe detuning behavior.

## 2 SYSTEM DESIGN OF SC LINAC

*2.1 Layout of SC linac*

Reference design of the SC proton linac system from 400MeV to 600MeV has been made. Figure 1 shows the schematic view of lattice structure. The SC linac is divided into two cavity groups because proton velocity increases as accelerating. The number of cells per cavity is 7. Each cryomodule unit consists of two cavities. The maximum electric surface peak field (Esp) of the cavities is 30MV/m which corresponds to the magnetic surface peak field Hsp of 525Oe. This criteria for the fixed maximum magnetic field limit is determined based on the experiences with multi-pacting condition of other SC cavity experiments. Average synchronous phase angle was set to be –30deg. The phase slip of the beam bunch in the 7-cell cavity was within ±16deg.

The lattice design has been performed by considering semi-equipartitioning condition of proton beam to reduce emittance growth. In this condition, the equipartitioning factor, $\gamma\varepsilon_{nx}\sigma_x /\varepsilon_{nz}\sigma_z$, (ratio between transverse and longitudinal values of emittance times phase-advance) was taken to be 0.8 rather than 1. Lengths of quadrupole magnets were determined from the limitation of Lorentz stripping of the negative hydrogen beams. Design criteria of stripping rate less than $10^{-8}$/m at bore radius (3cm) is adopted for the magnetic field gradient with 10% margin. Table 1 summarizes the design parameters. Quadrupole magnet length and distance between magnets are 45 cm. The Esp values are adjusted to achieve smooth phase advance between groups. Total number of the cryomodules is 15 with a total length of 69m.

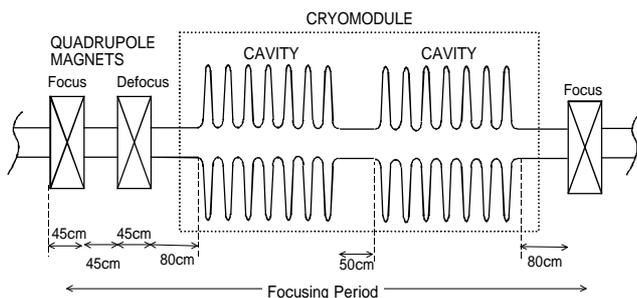

Fig. 1 Bloak diagram of lattice structure for SC linac

Table 1 Design parameters for the SC linac, $H_{sp}=525$ ($O_e$)

| β | L cm | $E_{sp}/E_{acc}$ | $H_{sp}/E_{acc}$ $O_e$/(MV/m) | K % | $E_{acc}$ MV/m | $E_{sp}$ MV/m |
|---|---|---|---|---|---|---|
| 0.73 | 5.62 | 3.05 | 53.0 | 2.9 | 9.91 | 30.2 |
| 0.77 | 5.94 | 2.85 | 50.1 | 2.6 | 10.47 | 29.9 |

## 2.2 Beam simulation

Beam simulation has been carried out with the modified PARMILA code using the parameters based on the semi- equipartitioning condition[4]. The RMS emittance growth rates in transverse and longitudinal direction are 5% and –2%, respectively. Effects of the RF phase and amplitude control error to the momentum spread were evaluated in the energy region between 400 to 600MeV. The phase and amplitude errors were introduced independently in the simulation assuming uniform distribution. Intrinsic energy spread of the injected beam at 400MeV was assumed to be ±0.2MeV. The 1000 cases of the calculations were carried out in the given error condition and the calculated averaged output energy was obtained as a histogram. The energy spread is then estimated using the standard deviation of the histogram. The values of standard deviations due to the ±1deg. phase error and ±1% amplitude error are ±0.23MeV and ±0.19MeV, respectively. The total energy spread was estimate to be ±0.36MeV which corresponds to $\Delta p/p=\pm 0.3\%$ by including the ±0.2MeV intrinsic error.

# 3 SC CAVITY DEVELOPMENT

The SC cavity development is continued on the basis of the design parameters for the JAERI original project (NSP)[5], of which accelerating frequency is 600MHz and number of cells in each cavity is 5. Essential differences with respect to fabrication method, electromagnetic performance and mechanical property to the cavity are not expected between two frequency schemes. In the development work, vertical tests of a single-cell cavity and 5-cell cavities ($\beta$=0.50 and 0.89) have been carried out.

## 3.1 Fabrication of 5-cell cavities

A 5-cell cavity of $\beta$=0.50 was fabricated in the KEK workshop. The $E_{sp}/E_{acc}$, $H_{sp}/E_{acc}$, $R/Q$ of the cavity are 4.67, 94.8Oe/(MV/m), and 77.1$\Omega$, respectively. Equator straight lengths at both end cells are adjusted to achieve flat electric field distribution on the beam axis. Unexpected troubles were encountered in the fabrication process resulting in the cavity structure with different cell lengths. Pretuning of this cavity was carried out. Maximum deviation of the peak field at each cell center was 37.5% before the pretuning. After the pretuning, the deviation was reduced within 0.7. Field flatness within 2.1% was finally achieved after the pretuning but cavity length became longer by about 6cm and frequency increased by about 16MHz.

A 5-cell cavity of $\beta$=0.89 was fabricated in Toshiba corporation. The $E_{sp}/E_{acc}$, $H_{sp}/E_{acc}$, $R/Q$ of the cavity are 2.04, 47.4Oe/(MV/m) and 443$\Omega$, respectively. Pretuning of the cavity was carried out. Maximum deviation of the peak field at each cell center was 23% before the pretuning. Field flatness was improved to 2% in the pretuning. Figure 3 shows the field distribution of the 5 cell cavity before and after pretuning.

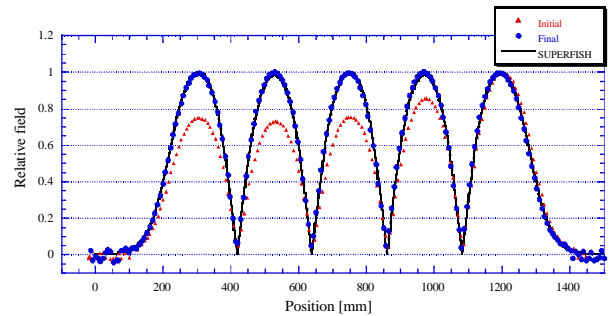

Fig. 2 Field distribution of the 5-cell cavity ($\beta$=0.89) before and after pretuning

## 3.2 Vertical test of 5-cell cavities

Surface treatments of the 5-cell niobium cavities ($\beta$=0.50 and $\beta$=0.89) were carried out with the same procedure for the single-cell cavity, i.e., barrel polishing (BP) and electro-polishing (EP). Average removal thicknesses in BP were 97$\mu$m and 89$\mu$m for $\beta$=0.50 and $\beta$=0.89 cavities, respectively. The EP processes were made to $\beta$=0.50 cavity twice and $\beta$=0.89 cavity three times with about 60$\mu$m and 90$\mu$m total removal thicknesses, respectively, before last vertical tests were carried out. In addition, the heat treatment at 750 C for 3hours and HPR (high pressure rinsing) for 1.5hours were done. In the vertical test, two kinds of curves between residual resistance and temperature of cooling down (Rs vs 1/T curve) and between quality factor and maximum surface electric field of the cavity (Q0 vs Esp curve) were obtained experimentally. Figure 3 shows the 5-cell cavity of $\beta$=0.89 mounted on the experimental set-up. The cavity was just taken out from the cryostat and covered with the frost.

Figure 4 shows the curve of the residual resistance as a function of 1/T. The surface resistance

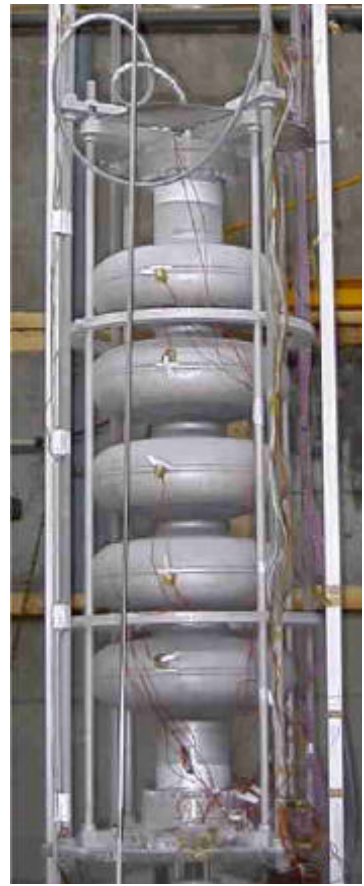

Fig. 3 A 5-cell cavity of $\beta$=0.89

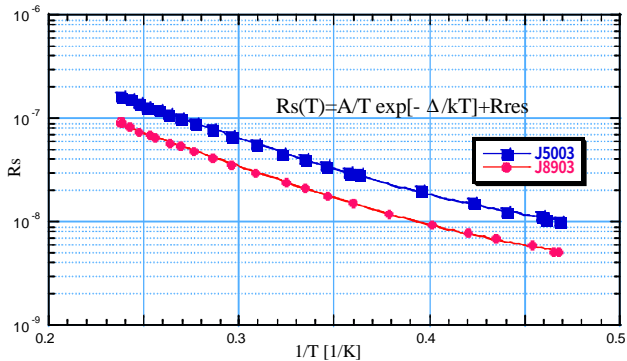

Fig.4 Residual resistance as a function of 1/T (inverse of cooling temperature)

values at 16MV/m for the β=0.50 and β=0.89 cavities are 10nΩ and 5nΩ at 2K, respectively.

Figure 5 shows the results with the vertical tests of the 5-cell cavities both for β=0.50 and β=0.89 at 4K and 2K. For the β=0.50 cavity experiments, maximum field strengths of 23 and 18.7MV/m were obtained in 2K and 4.2K measurements, respectively. The field was quenched at 2K and limited by the capacity of the RF power supply at 4K. Quality factors Q were reasonable at low field strength ($2 \times 10^{10}$ and $1 \times 10^9$ at 2K and 4.2K, respectively), but were degraded as field increased.

For the β=0.89 experiments, maximum field strengths of 23 and 31MV/m were obtained in 2.1K and 4.2K measurements, respectively. The fields were limited by thermal quench at 2K and the capacity of RF power supply at 4K due to a field emission. Good quality factors Q of $5 \times 10^{11}$ and $2 \times 10^9$ were obtained at 2K and 4K, respectively.

The field strengths exceeded design values of 16MV/m for original 600MHz cavity. These performances, however, were not good compared with the single-cell cavities which reached constantly to the values more than 40MV/m[5]. The reasons for these results are considered due to the cavity deformation in the pretuning for β=0.50 cavity and insufficient surface treatment both for β=0.50 and β=0.89. Further studies will be performed to improve the performances to meet the final requirement with the Esp value of more than 30MV/m.

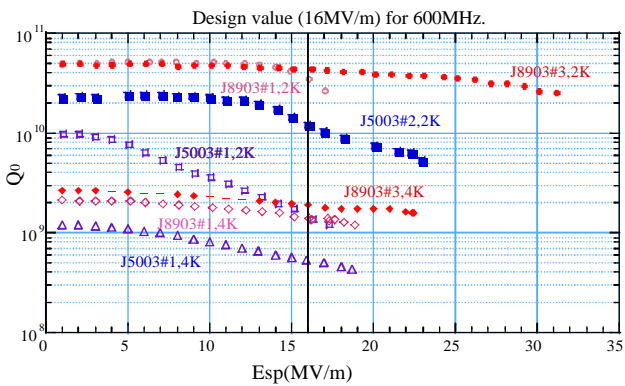

Fig.5 Vertical test results of the 5-cell cavities at 2K and 4K. Tests were done twice and three times for β=0.50 and β=0.89 cavity, respectively.

## 4 DYNAMIC ANALYSIS OF LORENTZ DETUNING

In the Joint Project, pulsed operation of the SC cavities is planned with repetition rate of 50Hz and beam pulse width of 0.5ms. Dynamic behavior of the Lorentz detuning due to pulsed operation is the most important issue for stable RF control of the cavities[6].

The dynamic analysis of the Lorentz detuning was performed with the finite element model code ABAQUS. The Lorentz force on the cavity wall was obtained from the electromagnetic field distribution which was calculated by the SUPERFISH code. The Lorentz detuning and the cavity field influence each other. To solve the dynamic Lorentz detuning and the dynamic cavity field simultaneously, Lorentz vibration model which describes dynamic behavior of the Lorentz detuning is established. A programming language of MATLAB/Simulink was used to solve the double differential equation. The model was applied to the simulation of the RF control successfully[7].

## 5 SUMMARY

System design of the SC proton linac has been carried out for the JAERI/KEK Joint Project. R&D work of the SC cavities for the high intensity proton linac has been progressing and promising results are accumulated. A model which describes the dynamic Lorentz detuning in the pulsed operation was established. Design of a prototype cryomodule, which includes two 5-cell cavities of β=0.60, is in progress. Cavity and cryomodule tests will be made early in 2001. Experiments of the RF control is planned using the prototype cryomodule.